\documentclass[usetikz]{cepcnote} 

\usepackage{listings}
\usepackage{color}
\usepackage{url}

\definecolor{mygreen}{rgb}{0,0.6,0}
\definecolor{mygray}{rgb}{0.5,0.5,0.5}
\definecolor{mymauve}{rgb}{0.58,0,0.82}

\skipbeforetitle{100pt}

\title{A proposed solution for analysis management in high energy physics}

\usepackage{authblk}
\renewcommand\Authands{, } 
\renewcommand\Affilfont{\itshape\small} 
\newcommand\my[1]{\emph{#1}}

\author[a,b]{Mingrui Zhao\footnote{The author was a student in Tsinghua University when this work started.}}

\affil[a]{Department of Nuclear Physics, China Institute of Atomic Energy, Beijing 102413, China}
\affil[b]{Center of High Energy Physics, Tsinghua University, Beijing 100084, China}

\abstracttext{
  This paper presents an architecture for the analysis management in high energy physics experiments. Some new concepts on data analysis are introduced. A protocol for organizing and operating an analysis is raised. A toolkit following this architecture is developed, which provides a solution of analysis management with both flexibility and reproducibility. A foreseen development of this toolkit is discussed.
}

\begin{document}

\section{Introduction}
A physics analysis should be well managed.
One of the most important aspects in analysis management is the analysis preservation.
Analysis preservation is essential in scientific research.
It is the responsibility of a researcher to clarify the details to produce the result from the raw data.
The details include the codes, data, analysis steps and other information needed to reproduce the results.
An analysis result which can not be reproduced is meaningless and should be regarded as being produced through ``magic''.
With the reproducibility of an analysis, knowledge is able to transfer between analyzers.
The young analyzer can learn the complete analysis steps from the analysis codes itself.
And if an analyzer leaves a group, his/her colleagues can pick up his/her work without barriers.

Another critical point in analysis management is its convenience.
The convenience and reproducibility should make a balance. A tool with high reproducibility but low convenience will force the 
users to bypass it. And thus the reproducibility will lose.

People often overlook the significance of analysis management.
They argue that management is just a bookkeeping work.
However, management of the analysis is not an easy job. Recording the analysis details to an extent that others can repeat the analysis require quite a lot manpower. It is the present situation that most analyses are not able to be reproduced. 

Why is analysis management difficult? One of the major sources of difficulty is from the preservation of data. Generally, data are sizable, and preserving the data using twice as many disks seems unwise. The codes and the data are usually lightly bound. The data are not generated intermediately after the codes are written or modified, leading the mismatch of analysis codes and results. If the data were generated from some code, it is usually impossible to read the code from the data.
Besides, the following problems or cases often occur in high energy physics analysis.
They make it harder to reproduce the analysis conveniently, as is pointed out in~\cite{atlas-talk}~and \cite{reana}.

\begin{itemize}
  \item Data and codes are frequently modified or moved. 
    For example, the shape of the fit functions can be adjusted.
    The upstream data may be modified and the whole analysis should rerun.
  \item Analyses in high energy physics are usually long-term,
    and they are usually highly complicated.
    Human beings are not reliable and analyzer often have terrible habits to organize the data and codes.
    The exact analysis step might be forgotten by the analyzer after a long time.
\end{itemize}


The importance of analysis management has been recognized within high energy physics community.
A conference preceding on data preservation\cite{data-preservation} is published in 2012.
The CERN analysis preservation portal(CAP)\cite{cern-analysis-preservation} is set up to prompt the systematic preservation of analysis for LHC experiments.
CAP provides a centralized platform on which scientists can document their analysis as early as the start of a new project.
CAP use REANA\cite{reana} as backend, 
which is a platform for reusable research data analyses, receiving join effort from CERN IT, SIS, DASPOS and DIANA-HEP.


In this paper, a proposal is raised to resolve the pain of analysis management. 
The proposal include an analysis architecture and a toolkit.
The recent development of the container techniques both conceptually inspires the designation and technically provides a tool to realize the designation, as will be talked about in details in the following sections.
The new management system will change the current behavior of analyzer as less as possible.
The management system allows the users to use whatever analysis tool they want.
A possible connection of this proposal with the CAP and REANA is also discussed.





\section{Conceptual designation}

\usetikzlibrary{graphs}
\usetikzlibrary{positioning}
\usetikzlibrary{decorations.markings}
\usetikzlibrary{decorations}
\usetikzlibrary{arrows}

\subsection{Workflow architecture}
\begin{figure}[htb]
     \begin{centering}
       \begin{tikzpicture}
         [ nonterminal/.style={
           rectangle,
         minimum size=6mm,
         very thick, draw=red!50!black!50,
         top color=white,
         bottom color=red!50!black!20, 
         font=\itshape
       }, 
         node distance=5mm, 
         terminal/.style={
           rectangle,minimum size=6mm,rounded corners=3mm, 
         very thick,draw=black!50,
         top color=white,bottom color=black!20, font=\ttfamily}
       ,>=stealth,thick,black!50,text=black, every new ->/.style={shorten >=1pt}
         ]

         \node (raw data) [terminal] {Raw data};
         \node (step1) [terminal, right=of raw data] {Step 1};
         \node (dataA) [terminal, right=of step1] {Data A};
         \node (step2) [terminal, right=of dataA] {Step 2};
         \node (dataB) [terminal, right=of step2] {Data B (Result)};
         \path (raw data) edge[->] (step1)
             (step1) edge[->] (dataA)
             (dataA) edge[->] (step2)
             (step2) edge[->] (dataB);
       \end{tikzpicture}
       \vfill
       \vspace*{5mm}
     \end{centering}
       \begin{centering}
         \begin{tikzpicture}
           [ nonterminal/.style={
             rectangle,
           minimum size=6mm,
           very thick, draw=red!50!black!50,
           top color=white,
           bottom color=red!50!black!20, 
           font=\itshape
         }, 
           node distance=5mm, 
           terminal/.style={
             rectangle,minimum size=6mm,rounded corners=3mm, 
           very thick,draw=black!50,
           top color=white,bottom color=black!20, font=\ttfamily}
         ,>=stealth,thick,black!50,text=black, every new ->/.style={shorten >=1pt}
           ]

           \node (raw data) [terminal] {Raw data};
           \node (step1) [terminal, above right=of raw data] {Step 1};
           \node (dataA) [terminal, right=of step1] {Data A};
           \node (step2) [terminal, below right=of raw data] {Step 2};
           \node (dataB) [terminal, right=of step2] {Data B};
           \node (step3) [terminal, right=of dataB] {Step 3};
           \node (dataC) [terminal, right=of step3] {Data C};
           \node (step4) [terminal, above right=of dataC] {Step 4};
           \node (dataD) [terminal, right=of step4] {Data D(Result)};
           \path (raw data) edge[->] (step1)
             (step1) edge[->] (dataA)
             (raw data) edge[->] (step2)
             (step2) edge[->] (dataB)
             (dataB) edge[->] (step3)
             (step3) edge[->] (dataC)
             (dataA) edge[->] (step4)
             (dataC) edge[->] (step4)
             (step4) edge[->] (dataD)
             ;
         \end{tikzpicture}

       \end{centering}
     \caption{Examples of analysis workflow.
     The upper one represents a simple analysis and the lower is a slightly complicated one.}
     \label{fig:1}
   \end{figure}
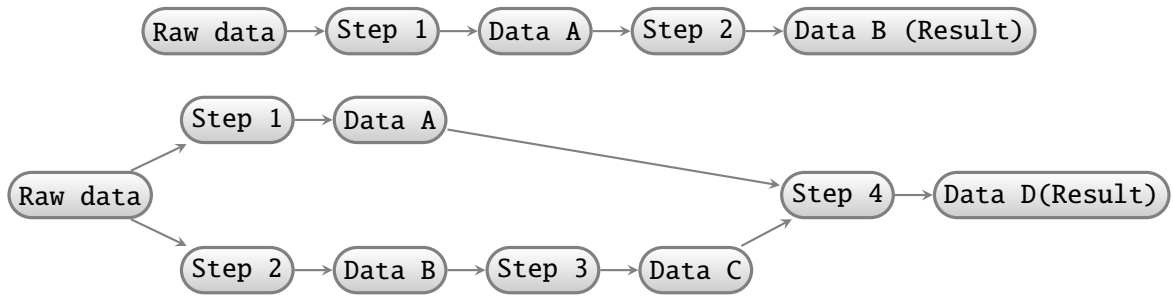

An analysis is always step by step.
Each step requires some inputs and generates some outputs.
If the inputs, outputs and steps are regarded as vertex,
and directed edges are linked from the inputs to steps, as well as from steps to the outputs.
The structure of an analysis naturally forms a directed acyclic graph, which can be called workflow.
Fig.~\ref{fig:1} shows examples of workflow.

   Workflow exists in all analysis. 
   The workflow of analysis contains fruitful information.
   It tells the procedure of the analysis,
   and it is also an invariant when the physical locations or names of the data or codes change.

   In the traditional way of analysis management,
   workflow is usually implicitly defined through the data location written in the codes, and it is usually ignored and seldom used.
   Workflow in analysis is also used in other domains like bioinformatics, medical imaging, astronomy, and chemistry.
   Workflow languages such as CWL\cite{cwl} are developed. 
   This explicit way of describing workflow has the advantage that the analysis procedure is transparent and directly readable to both the machine and the analyzer, clarifying the exact steps to do the analysis. So that the machine can run the analysis according to it, and analyzer can obtain knowledge of the analysis from it.

   Additionally, there are two observations: 
   \begin{itemize}
   \item A step in workflow does not necessarily know the whole workflow. It need only to know its predecessors to execute this specific step. 
   \item The step can be abstract, i.e. only the topological structure of the analysis matters.
   \end{itemize}


   Although workflow is widely used in management tool, it is seldom realized that the key function of workflow in analysis preservation is providing a tight binding between codes and data.

   The typical workflow concept has some slight differences with the one in this paper.
   In the workflow languages, they describe the exact steps to do the analysis.
   In this architecture, I choose to describe the workflow in a half-explicit way.
   It means that the workflow is not defined implicitly as described above.
   And it is not also defined explicitly in a sheet using some workflow language.
   It is defined through the relationship between steps. 
   The exact name or directory of an analysis step is unimportant. 
   When they are changing, the analysis keeps invariant.
   The different methods to describe the workflow are obviously equivalent. However, the half-explicit way make the problem clear and is more suitable to my designation, as will be shown in the following sections.



\begin{figure}[tb]
     \begin{centering}
       \begin{tikzpicture}
         [ nonterminal/.style={
           rectangle,
         minimum size=6mm,
         very thick, draw=red!50!black!50,
         top color=white,
         bottom color=red!50!black!20, 
         font=\itshape
       }, 
         node distance=5mm, 
         terminal/.style={
           rectangle,minimum size=6mm,rounded corners=3mm, 
         very thick,draw=black!50,
         top color=white,bottom color=black!20, font=\ttfamily}
       ,>=stealth,thick,black!50,text=black, every new ->/.style={shorten >=1pt}
         ]

         \node (raw data) [terminal] {Raw data};
         \node (task1) [terminal, right=of raw data] {Task 1};
         \node (dataA) [terminal, right=of task1] {Data A};
         \node (task2) [terminal, right=of dataA] {Task 2};
         \node (dataB) [terminal, right=of task2] {Data B (Result)};
         \node (algorithm1) [terminal, above=of task1] {Algorithm 1};
         \node (algorithm2) [terminal, above=of task2] {Algorithm 2};
         \path (raw data) edge[->] (task1)
             (task1) edge[->] (dataA)
             (dataA) edge[->] (task2)
             (task2) edge[->] (dataB)
             (algorithm1) edge[->] (task1)
             (algorithm2) edge[->] (task2);
       \end{tikzpicture}
       \vfill
       \vspace*{5mm}
     \end{centering}

     \caption{Example of workflow with separated algorithm and task.}
     \label{fig:2}
   \end{figure}
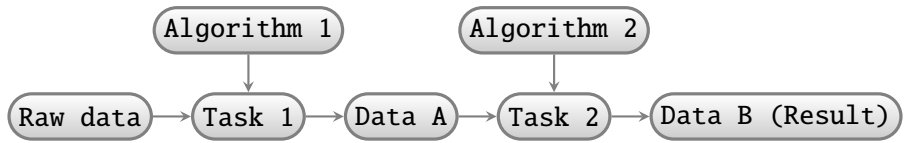

   Some of the analysis steps could be quite similar. They are better to be reused.
   I would like to separate the common part and the specialized part of an analysis step as ``algorithm'' and ``task''.
   Loosely speaking, an ``algorithm'' contains the environment definition and the code template.
   An ``algorithm'' is will be used to run a ``task'' and the ``algorithm'' can be shared by ``tasks'', while a ``task'' contains parameters. 
   The rigorous definition of ``algorithm'' and ``task'' will be in the next section. And they will be written in \my{italics} to clear up the possible ambiguities.
  The workflow of an possible analysis after the separation of ``algorithm'' and ``task'' is shown in Fig~\ref{fig:2}.

   It is also possible to refine the analysis architecture further. A task is always followed by some data.
   It is convenient to merge the ``data'' and ``task'' into one object. And the merge makes the binding of codes and data even tighter.
   
\begin{figure}[tb]
     \begin{centering}
       \begin{tikzpicture}
         [ nonterminal/.style={
           rectangle,
         minimum size=6mm,
         very thick, draw=red!50!black!50,
         top color=white,
         bottom color=red!50!black!20, 
         font=\itshape
       }, 
         node distance=5mm, 
         terminal/.style={
           rectangle,minimum size=6mm,rounded corners=3mm, 
         very thick,draw=black!50,
         top color=white,bottom color=black!20, font=\ttfamily}
       ,>=stealth,thick,black!50,text=black, every new ->/.style={shorten >=1pt}
         ]

         \node (raw data) [terminal] {Raw data task};
         \node (task1) [terminal, right=of raw data] {Task 1};
         \node (task2) [terminal, right=2cm of task1] {Task 2};
         \node (algorithm1) [terminal, above=of task1] {Algorithm 1};
         \node (algorithm2) [terminal, above=of task2] {Algorithm 2};
         \path (raw data) edge[->] (task1)
             (task1) edge[->] (task2)
             (algorithm1) edge[->] (task1)
             (algorithm2) edge[->] (task2);
       \end{tikzpicture}
       \vfill
     \end{centering}

     \caption{Example of workflow with combination of task and data.}
     \label{fig:3}
   \end{figure}
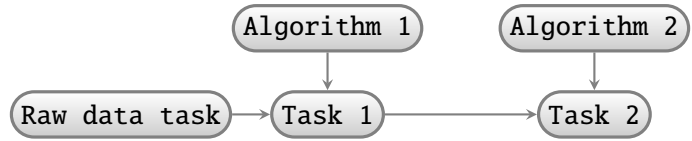

   The final architecture is illustrated in Fig.~\ref{fig:3}.


   \subsection{Preservation and ``impression''}
The code and the data in the above designation are still loosely bounded, since the modification on the codes would not impact the data immediately. There exists a conflict between the requirement of modification of codes and the preservation of the corresponding data. In another word, a difficulty in the analysis management is the balance of its flexibility and reproducibility. 
   To solve the difficulty, the concept of ``impression'' is introduced.
      \subsubsection{Abstract and concrete layers}
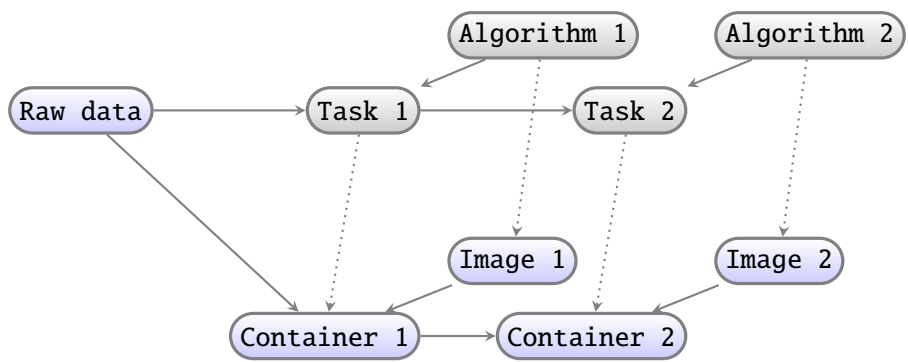
\begin{figure}[tb]
     \begin{centering}
       \begin{tikzpicture}
         [ nonterminal/.style={
           rectangle,
         minimum size=6mm,
         very thick, draw=red!50!black!50,
         top color=white,
         bottom color=red!50!black!20, 
         font=\itshape
       }, 
         node distance=5mm, 
         terminal/.style={
           rectangle,minimum size=6mm,rounded corners=3mm, 
         very thick,draw=black!50,
         top color=white,bottom color=black!20, font=\ttfamily}
       ,concrete/.style={
         rectangle,minimum size=6mm,rounded corners=3mm, 
         very thick,draw=black!50,
         top color=white,bottom color=blue!20, font=\ttfamily}
       ,>=stealth,thick,black!50,text=black, every new ->/.style={shorten >=1pt}
         ]

         \node (raw data) [concrete] {Raw data};
         \node (task1) [terminal, right=2cm of raw data] {Task 1};
         \node (task2) [terminal, right=2cm of task1] {Task 2};
         \node (algorithm1) [terminal, above right=of task1] {Algorithm 1};
         \node (algorithm2) [terminal, above right=of task2] {Algorithm 2};

         \node (image1) [concrete, below right=of task1, yshift=-1cm] {Image 1};
         \node (image2) [concrete, below right=of task2, yshift=-1cm] {Image 2};
         \node (container1) [concrete, below left=of image1] {Container 1};
         \node (container2) [concrete, below left=of image2] {Container 2};

         \path
             (raw data) edge[->] (task1)
             (task1) edge[->] (task2)
             (algorithm1) edge[->] (task1)
             (algorithm2) edge[->] (task2)

            (task1) edge[->,dotted] (container1)
             (task2) edge[->,dotted] (container2)
             (algorithm1) edge[->,dotted] (image1)
             (algorithm2) edge[->,dotted] (image2)

            (raw data) edge[->] (container1)
             (container1) edge[->] (container2)
             (image1) edge[->] (container1)
             (image2) edge[->] (container2);

       \end{tikzpicture}
       \vfill
     \end{centering}
     \caption{Example of abstract and concrete layer.}
     \label{fig:4}
   \end{figure}

Noticing the conflicts between different requirements discussed above, the analysis can be split into abstract layer and concrete layer according to the requirement of flexibility or reproducibility. To be specific, the abstract layer of an analysis is what the author write or operate. 
   The concrete layer is a copy of the abstract layer, while the machine-useless information is stripped.
   The machine-useless information is only for the convenience of analyzer, such as the path of putting a task.
   The concrete layer can really be run and that is why it is called ``concrete''.
   The corresponding entity of ``algorithm'' and ``task'' in concrete layer are ``image'' and ``container'',
   whose concepts are borrowed from Docker\cite{docker}. 
   They can be executed to generate data. 
   After the separation of the abstract layer and the concrete layer,
   the topological structure of an analysis may look like Fig.~\ref{fig:4}.

   Since the abstract layer of the analysis contains only codes and some metadata, it has quite small size. They can be easily shared between analyzer to reuse and review.
   And the abstract layer can also be uploaded to a central portal for centralized management.


   \subsubsection{``Impression''}
The reader may be confused about how such a splitting can balance the flexibility with reproducibility. In fact, it should move further to prompt ``image'' and ``container'' to ``impression''.
   An ``impression'' is a version of an ``image'' or an ``container''. It is able to generate data uniquely. And it is tightly bounded with the ``task'' and ``algorithm'' in the abstract layer. Therefore, a tight binding of codes and data, using an intermediate ``impression'', is established.
   The abstract layer of analysis is allowed to be arbitrary modified. Once it is modified, the link between the abstract layer and the original ``impression'' will break immediately, i.e. the data is sensitive to the codes. Therefore, the objects 
   necessary to preserve are only codes and corresponding ``impressions'', which are all have small size.


   For example, as shown in Fig.~\ref{fig:1}, assume the original version of the analysis has four impressions,
   ``Image 1.1'', ``Image 2.1'', ``Container 1.1'', and ``Container 2.1''.
   After the modification of ``Algorithm 2'', 
   a new impression ``Image 2.2'' is created. At the same time, the ``Task 2'' will also 
   generate a new impression ``Container 2.2''.
   Although the contents ``Task 2'' does not change, the upstream dependence of ``Task 2'' changed.
   The workflow of the concrete part analysis keep changing when the analysis is on going.
   That is why a half-explictly way of constructing workflow is more suitable for the case.

   The abstract layer contains all the history of impression. Analysis can be reverted to a specific version 
   if necessary.

   \begin{figure}[tb]
     \begin{centering}
       \begin{tikzpicture}
         [ nonterminal/.style={
           rectangle,
         minimum size=6mm,
         very thick, draw=red!50!black!50,
         top color=white,
         bottom color=red!50!black!20, 
         font=\itshape
       }, 
         node distance=5mm, 
         terminal/.style={
           rectangle,minimum size=6mm,rounded corners=3mm, 
         very thick,draw=black!50,
         top color=white,bottom color=black!20, font=\ttfamily}
       ,concrete/.style={
         rectangle,minimum size=6mm,rounded corners=3mm, 
         very thick,draw=black!50,
         top color=white,bottom color=blue!20, font=\ttfamily\tiny}
       ,>=stealth,thick,black!50,text=black, every new ->/.style={shorten >=1pt}
         ]

         \node (raw data) [concrete] {Raw data};
         \node (task1) [terminal, right=4cm of raw data] {Task 1};
         \node (task2) [terminal, right=4cm of task1] {Task 2};
         \node (algorithm1) [terminal, above right=of task1] {Algorithm 1};
         \node (algorithm2) [terminal, above right=of task2] {Algorithm 2};

         \node (image1-1) [concrete, below right=of task1, yshift=-1cm] {Image 1.1};
         \node (image2-1) [concrete, below right=of task2, yshift=-1cm] {Image 2.1};
         \node (image2-2) [concrete, below=of image2-1] {Image 2.2};
         \node (container1-1) [concrete, below left=of image1-1, yshift=-2cm] {Container 1.1};
         \node (container2-1) [concrete, below left=of image2-1, yshift=-2cm] {Container 2.1};
         \node (container2-2) [concrete, below=of container2-1] {Container 2.2};

         \path
             (raw data) edge[->] (task1)
             (task1) edge[->] (task2)
             (algorithm1) edge[->] (task1)
             (algorithm2) edge[->] (task2)

            (task1) edge[->,dotted] (container1-1)
             (task2) edge[->,dotted] (container2-1)
             (algorithm1) edge[->,dotted] (image1-1)
             (algorithm2) edge[->,dotted] (image2-1)

            (raw data) edge[->] (container1-1)
             (container1-1) edge[->] (container2-1)
             (container1-1) edge[->] (container2-2)
             (image1-1) edge[->] (container1-1)
             (image2-2) edge[->] (container2-2)
             (image2-1) edge[->] (container2-1);

         \node (container2-1) [concrete, below left=of image2-1, yshift=-2cm] {Container 2.1};

     \end{tikzpicture}
       \vfill
   \end{centering}
     \caption{Example of abstract and concrete layer. The container and images are all impressions.}
     \label{fig:5}
 \end{figure}
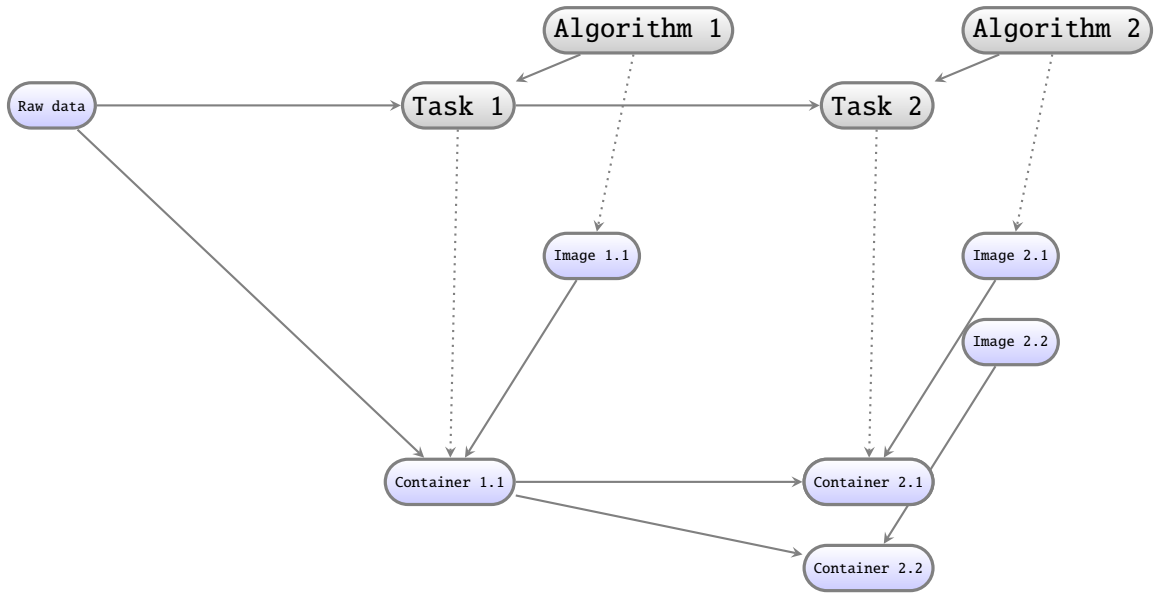

 \subsection{Runner}
 ``Impression'' is designed to determine the result. 
Conceptually, there should be a standalone runner to run the impression. The runner can be on a local machine and on a remote machine. The Analyzer sends ``impression'' to the runner. The runner executes the ``impression'' and generates result. And then the analyzer gets the results from the runner.
 The ``containers'' with tiny resource consumption can be run on 
   the personal computer.
   For the ``containers'' with huge resource consumption, the analyzer
   can choose to transfer them to computing grids or powerful computing machines to compute. 
   The computed containers can be transferred back to personal computer to view the results or to satisfy the dependence of
   executing the next container on the personal computer.

 \subsection{Status}
Each ``task'', ``algorithm'' or ``impression'' can attach an status. 
 The status is determined so that runner can manage the job sequence and analyzer can monitor the procedure of the analysis.
   \subsubsection{Status of ``image'' and ``container''}
   Since an ``impression'' is immutable, it is not difficult to determine the status of an impression.
   If an ``impression'' is executed, the status of it will be ``built'' or ``done''
   for ``image'' and ``container'' separately. The on going status is ``building'' or ``running''.
   If it happened to be failed, a ``failed'' status is assigned.

   \subsubsection{Status of ``task'' and ``algorithm''}
   The status of algorithm and task can be ``impressed'' or ``new''.
   One condition of ``impressed'' status is that the 
   contents of an algorithm or task is the same as that of the corresponding impression.
   And the other condition is that dependences of the two parts also match.
   Otherwise, the status is assigned as ``new''.
   
   \subsection{Global link}
  Having the ``impression'' in hand, it is natural to link all the analysis in a collaboration together. An analyzer will not ask his/her colleague for data. The change of the upstream analysis will also be detected by the analyzer.

\section{Protocol}

\subsection{Introduction}
This protocol is designed to provide a standard way to organize
the analysis in the high energy physics. If the analysis is organized in a standard way, the toolkit developer can develop software helping analyzer to manage it. The designation aims to validate conceptual designation of the analysis management system. The toolkit developer can benefit from a standard way of organization and develop tool to make the operation easier.


\subsection{Organization}

\subsubsection{Chern repository}
All the analysis codes and metadata should be put in a directory. This directory is called a Chern repository. In the Chern repository, there exists a lot of subdirectories. Some of the directories contain hidden folder called ``.chern''. Directory with the hidden ``.chern'' is called an \my{object}. The root directory of the repository is also an \my{object}. And it should be addressed that an \my{object} is a directory in the file system.

There are four types of \emph{object}. They are \emph{algorithm}, \emph{task}, \emph{directory} and \emph{project}. Each object contains a file ``README.md'' used for documentation purpose. And a \emph{object} also has a file named ``.chern/config.json'' serving as the configuration file of the \my{object}. The type of the object(\my{algorithm}, \my{task}, \my{directory} and \my{project}) is written in the configuration file of the \my{object}. In addition to the files common in all kinds of \my{object}s. The different types of \my{object} have different contents.
\begin{itemize}
\item A \emph{directory} is just a folder to contain other sub\emph{object}s. The \emph{object} in a \emph{directory} can be any type except \my{directory}.
\item A \emph{project} is exactly the same as the \emph{directory} except that it is the root of the whole analysis repository.
\item An \emph{algorithm} contains the code templates and environment specification files(Dockerfile for temporatory), provided by collaboration. There is a folder called ``.chern/impressions'' containing all the impressions corresponding to the \my{algorithm}.
\item In a \my{task}, there should be a file ``parameter.json'' containing the parameters of the task. There is also a folder called ``.chern/impressions'' as \my{algorithm}.
\end{itemize}

The connection between the \my{task}s and \my{algorithm}s, i.e. workflow, is specified in the configuration files of the \my{task}s and \my{algorithm}s. In the configuration of a \my{task}, the \my{algorithm} and \my{task}s that this \my{task} depends on are recorded, in the form of the relative path to the \my{project}.  The depending \my{task} has also an ``alias'' recorded in this configuration file. In the configuration file of a \my{task} or \my{algorithm}, the \my{task}s it depending are recorded.

Other files or directories that are not organized as above statements, do not belong to the repository and should be ignored.


\begin{figure}[tb]
\begin{center}
\includegraphics[width=0.35\textwidth]{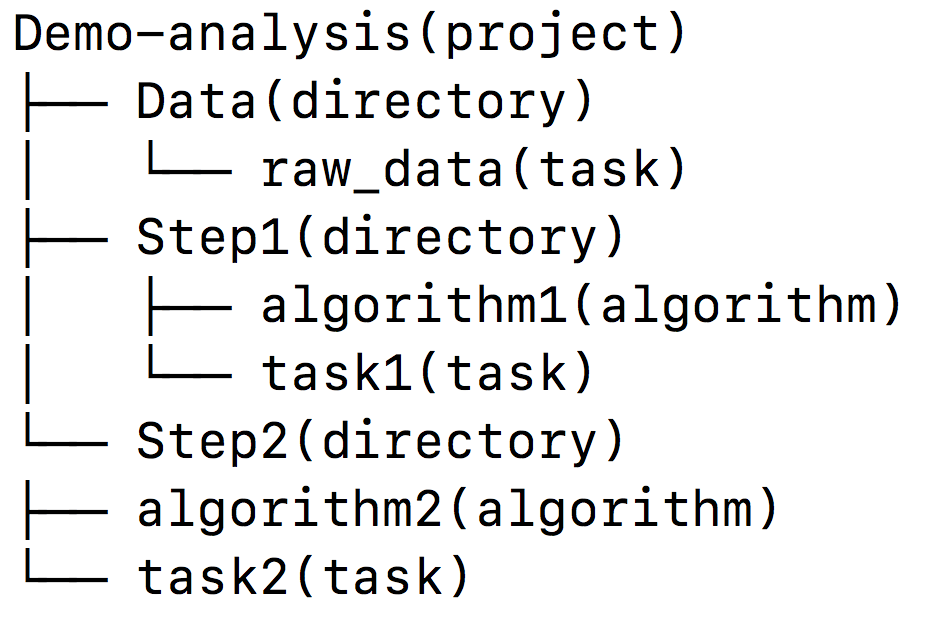}
\end{center}

\caption{A possible analysis organization.}
\label{fig:ana}
\end{figure}

\subsubsection{\my{Impression}} 
An \my{impression} is a version of a \my{task} or \my{algorithm}. It has a unique id and is physically stored in the ``.chern/impressions/(ID)'' under the \my{task} or \my{algorithm}, where the (ID) refer to the id of the \my{impression}. Under the ``.chern/impressions/(ID)'', there exists a file named ``config.json'', called the configuration file of the \my{impression}. And there is also a folder named ``contents'' to store the contents of the corresponding \my{task} or \my{algorithm}.
In the configuration file, the file tree of contents is written. It also has the historical \my{impression} id and the \my{impression}s that this \my{impression} depends on.

The above information stored in a impression is enough for running the impression anywhere.

     \subsubsection{Parameter transfer and I/O map}
     The parameters defined in a \my{task} should be able to be transfered to the corresponding \my{algorithm}.
     Let us take an ROOT or C++ application as example. When the analyzer writes a template code in ROOT for an \my{algorithm},
     he/she could include a header file called ``arguments''.
     After including the header, he/she can use ``parameters[(PARAMETER\_NAME)]'' as a std::string type value in the template code,
     where (PARAMETER\_NAME) represents the parameter defined in the \my{task}.

     Dealing with the inputs and outputs is similar. 
     As discussed above, each predecessor of the \my{task} has an alias. After including the header, analyzer can use ``folder[(ALIAS)]'' as a std::string type value. This value means the storage location of the data of the predecessor \my{task}. The output directory to store the result is just ``folder[``output'']''.
     
The runner should generate the header file ``arguments''.


     Other programming languages
     can be supported in a similar way. However, this do not mean that the standard transfer header should be created for all the programming language that we can image. Creating ``header file'' for script language like bash, tcsh or python, and using them to start unsupported  language is feasible.
     
\subsubsection{Pure data \my{task}}
It is a common case that the starting point of an data analysis is external data. However, as the ``data'' can only exists in a ``task'', there is no place to put the external data.
This can be solved in the following way. The Analyzer creates a file named ``data.json'' under a \my{task} and pretend that it can be executed and generate data. The md5 of the data expected to generate are recorded in ``data.json''. The ``data.json'' can not actually be run, and the analyzer should manually feed the external data to the runner, pretending that the data is generated by the \my{task}.

\subsection{Operations}
The operation applied to the repository should be restricted in order to keep the structure of the operation.
The following operations are allowed to use and should be regarded as the atomic operations. 
\begin{itemize}
\item Creating repository: making an empty folder and creating file ``.chern/config.json'' and ``.chern/project.json'' under it. Writing the \my{object} type of ``project'' into file ``.chern/config.json''.
\item Creating \emph{directory}: making an empty folder and creating file ``.chern/config.json'' under it. Writing the \my{object} type of ``directory'' into file ``.chern/config.json''. 
\item Creating \emph{algorithm}: making an empty folder and creating file ``.chern/config.json'' under it. Writing the \my{object} type of ``algorithm'' into file ``.chern/config.json''.
\item Creating \emph{task}: \emph{algorithm}: make an empty folder and creating file ``.chern/config.json'' under it. Writing the \my{object} type of ``task'' into file ``.chern/config.json''.
\item Link an \my{algorithm} and a \my{task}: adding the \my{task} to the successor list of the \my{algorithm} in the configuration file of the \my{algorithm}. Adding the \my{algorithm} to the predecessor list of the \my{task}.
\item Unlink and \my{algorithm} and a \my{task}: removing the \my{task} from the successor list of \my{algorithm}. Removing the \my{algorithm} from the predecessor list of the \my{task}.
\item Link a \emph{task} to a second \emph{task}: adding the second \my{task} to the successor list of the first \my{task}. Adding the first \my{task} to the predecessor list.
\item Unlink the relation of a \emph{task} to a second \emph{task}: 
removing the second \my{task} of the successor list of the first \my{task}. Removing the first \my{task} from the predecessor list of the second \my{task}. Removing the alias.
\item Moving \emph{object}: moving the directory. Since the \my{object} recorded in its predecessors and successors in the form of the path relative to the root of the whole repository. The configuration files of its predecessors and successors should be modified to use the new relative path.
\item Copying \emph{object}: copying the directory. Unlink the predecessors and the successors.
\item Removing \emph{object}: unlink the predecessors and the successors. Removing the directory.
\item Modify ``README.md'': just editing ``README.md''.
\item Creating \my{impression}: Once an impression is required to be created for an \my{algorithm} or a \my{task}, firstly, the \my{impression} of the preceding task or algorithm is checked to be available. After that, the contents of the task or algorithm and the preceding impression define the required \my{impression}.
\end{itemize}

\section{Technical designation and prototype}

A prototype toolkit named ``Chern'' is designed to realize the conceptual designation.
The github repository of the toolkit is on \url{https://github.com/hepChern}. And the documents for user can be found in \url{http://chern.redthedocs.io/en/latest}.
A few examples including some toy examples and some real analyses can be found on github.

\subsection{Frontend}
\begin{figure}[tb]
     \begin{center}
     \includegraphics[width=0.7\textwidth]{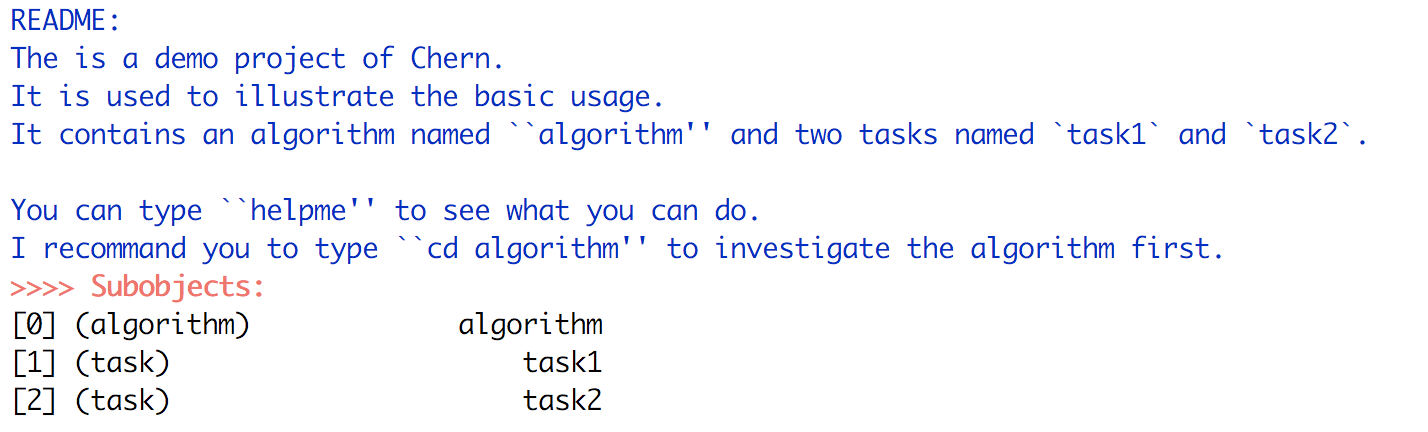}
     \end{center}
     \vspace*{5mm}
     \begin{center}
     \includegraphics[width=0.7\textwidth]{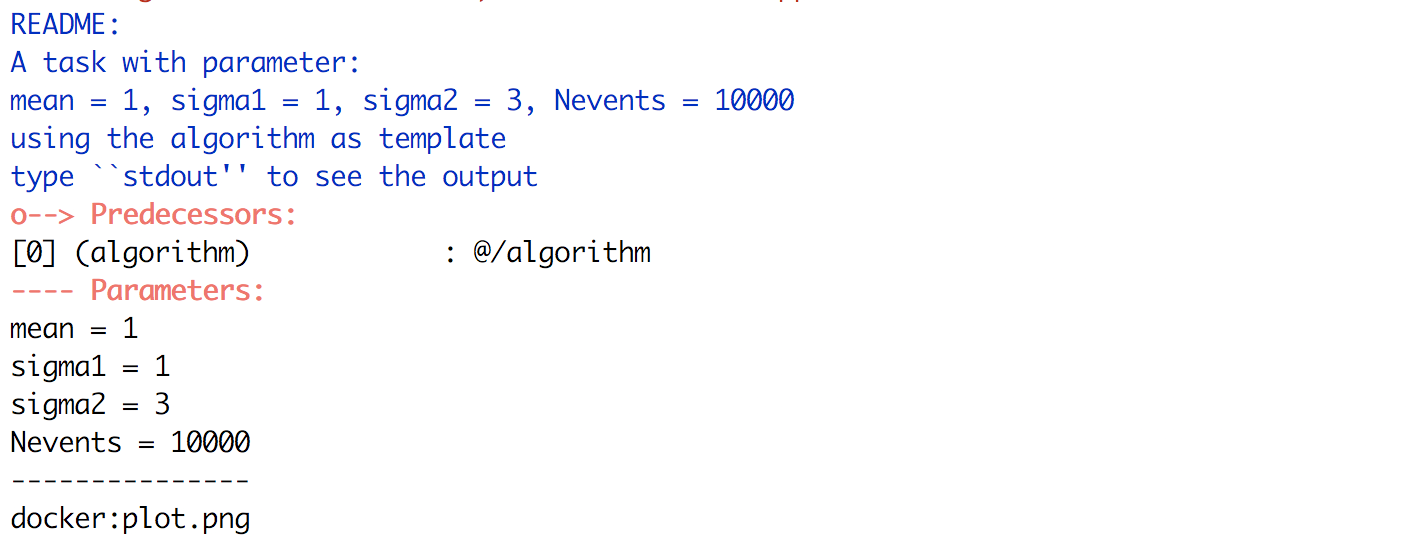}
     \end{center}

     \caption{The shown contents when the user type ``ls'' under a folder or a task.}
     \label{fig:ls}
     \end{figure}

A toolkit should develop commands to realize the operation defined in the above section and their combination helping the analyzer to operate the Chern repository conveniently.
I choose to use IPython notebook to hide the details of managing Chern repository as much as possible. The commands such as ``mv'', ``cd'', ``ls'', ``cp'' and many others are redefined.
For example, ``ls'' will not only show the contents of a directory, but also additional information like status, predecessors and so on as in Fig~\ref{fig:ls}.
Other helpful commands such as ``mktask'' to create a new \my{task}, ``submit'' to submit the current \my{object} to the runner, and other helpful commands are developed. 






\subsection{Chern Machine}

\subsubsection{Execution of \my{impression}}
     \my{Impression} should be able to be executed on any machine to generate the exact same results.
     In the past, to achieve such a requirement, a virtual machine is needed.
     Nowadays, the development of container technique makes such a requirement much easier to achieve.
     The runner will mount the storage of container and link the containers. The ``arguments'' file is created by runner at this time.
     

\subsubsection{Communication}
The communication protocols between the Chern repository and the runner, and between runners are not defined in the protocol above, since different communication methods are in principle equivalent and it is better to support many of them. In the current prototype, I only use http for the communication between Chern repository and runner. The exchange of impression and results between runners on different machines is not realized in the current prototype.  


\section{Discussion and Future development}
\subsection{Use cases and advantages}
In the following cases, using the proposed Chern architecture will be convenience.
They often occur in the analysis of high energy physics.
     \begin{itemize}
         \item Using the management system, analyzers can directly read the analysis steps from the repository.
         \item Analysis steps are preserved and for every data. The detailed steps to generate data are clear.
\item The name and location of the directory are not useful for backend to run the analysis.
     Therefore, they can be arbitrary modified without changing any result. When adding new parallel steps to analysis, this feature is especially useful.
     \item Similar analysis can easily be constructed. An analyzer can just copy a Chern repository from others and perform some slight tunning
     to create a new analysis.
     \end{itemize}

     \subsection{Future development}
%

     \subsubsection{Time machine}
     Although the full history is recorded in the Chern repository.
     The time machine function is not realized in the prototype. Reverting the analysis is often useful and this function will be added in the future version of the toolkit.
     
     \subsubsection{Monitoring}
     A job monitoring system can be developed to manage the job status and disk usage.

     \subsubsection{Frontend}
     The current frontend of the toolkit is based on IPython notebook. But the architecture is in fact independent of the analysis. 
     The frontend can also be web application or GUI application.
     They might be more user-friendly to the analyzers.

     \subsubsection{Backend}
     Technically, the current prototype of Chern only naively run, while RENAN support a lot of advanced techniques. REANA uses container technologies(Docker) and aims at supporting several different declarative workflow engines(Yadage, CWL), 
     several different computing cloud back-ends for job execution(Kubertenetes, HTCondor) 
     and several different storage back-ends for data sharing(EOS, Ceph).
     Using REANA as backend could make Chern more powerful.
     However, an interface between them require a lot effort.

\section{Conclusion}
In this paper, an architecture to manage an analysis is proposed.
With the new architecture, the workflow is constructed in an apparent way without losing convenience.
The requirement of preservation is also satisfied.
A novel toolkit is provided for analyzers.
So that they can do the analysis as what they do in the daily life today.

The future development of the architecture and the toolkit is also discussed.
The standard, time machine function, the new frontend and the new backend are also proposed.

\section{Acknowledgements}
I would like to express my acknowledgements to Sidan Chen, Mengzhen Wang, Li Xu, and Zhenwei Yang who help to polish this paper.

\bibliographystyle{plain}
\bibliography{chern}
\end{document}